\documentclass[12pt,a4paper]{article}
\usepackage{lineno,hyperref}
\usepackage[latin5]{inputenc}
\usepackage{amsmath}
\usepackage{amssymb}
\usepackage{graphicx}
\usepackage{subfigure}
\usepackage{caption}
\usepackage{float}
\usepackage{mathtools}
\usepackage{color}
\usepackage{mathrsfs}
\usepackage[]{geometry}
\usepackage{subfigure}

\newtheorem{theorem}{Theorem}
\title{Exact Solutions to Some Conformable Time Fractional Equations in Benjamin-Bona-Mahony Family}
\author{Alper Korkmaz\thanks{alperkorkmaz7@gmail.com} \\
{\scriptsize Çankırı Karatekin University, Department of Mathematics, 18200, Çankırı, Turkey.}} 
\begin{document}
\maketitle
\begin{abstract}

\noindent
The conformable time fractional forms of some partial differential equations are solved in the study. The existence of chain rule and the derivative of composite function enable the equations to be reduced to some ordinary differential equations by using some particular wave transformations. The modified Kudryashov method implemented to derive the exact solutions for the Benjamin-Bona Mahony (BBM), the symmetric BBM and the equal width (EW) equations in the conformable fractional time derivative forms. The obtained explicit solutions are illustrated for some time fractional orders in the interval $(0,1]$. 
\end{abstract}
\section{Introduction}

\noindent
The time fractional forms of the classical well-known partial differential equations in nonlinear forms are extension of the definitions of these equations. Even though there exist several fractional derivative definitions in the literature, the conformable one is studied throughout the paper. 

\noindent
The BBM equation is defined to describe the formation of some particular undular bore by a long wave in shallow water\cite{peregrine1}. The derivation of the BBM equation dates back to the wave phenomena in the water and the ion acoustic waves occurring in plasma\cite{dag1}. An analytical solution for an initial boundary value problem with some particular complementary data is suggested by Benjamin et al.\cite{benjamin1}. The Lagrangian density and the interaction of two solitary waves for the BBM equation are developed by Morrison et al.\cite{morrison1}. Wazwaz and Triki\cite{wazwaz1} propose some bright-type soliton solutions of the time dependent coefficient BBM equation with a simple dumping term. The Adomian decomposition method is another method to design some of the exact solitary wave solutions of the generalized form of the BBM equation\cite{kaya1}. Besides the analytical and exact solutions of the BBM equation, many numerical techniques from different families are developed and implemented for the numerical solutions to various evolution problems for the BBM equation\cite{korkmaz1,korkmaz2,gardner1}.  

\noindent
The symmetric BBM equation is defined by Seyler and Fenstermacher\cite{seyler1} to describe ion acoustic and space charge waves in the weakly nonlinear sense. Hyperbolic secant type solitary wave solutions and several invariants are also reported in the same work. Some exact solutions including powers of some hyperbolic and trigonometric functions of the symmetric BBM equation are developed by application of $(G'/G)$-expansion method\cite{abazari1}. Dereli's\cite{dereli1} study deals with meshless kernel method of lines to solve some initial boundary value problems covering motion of single solitary wave and the collusion two solitary waves with positive heights.

\noindent
Morrison et al.\cite{morrison1} obtain the equal width (EW) equation with its solitary wave solution for particular choices of some parameters. Owing to the linear relation between the EW and the BBM equations, one can conclude that the EW equation also has only three polynomial conservation laws like the BBM\cite{morrison1,olver1}. Some numerical approximations to the problems constructed on the EW equation can be listed as the method of lines with meshless kernel\cite{dereli2}, lumped Galerkin \cite{esen1} and the differential quadrature method\cite{saka1}. 

\noindent
So far, various methods from various expansion methods to hyperbolic ansatzes, from simple equation to first integral methods have been derived for the solutions of the class of nonlinear partial differential equations. Afterwards, the usage of those methods have been extended to solve nonlinear fractional-type partial differential equations\cite{bekir1,zhang1,ozkan1}. An alternative to those methods, Kudryashov modifies an old method to determine the exact solutions of the Fisher and a seven-order equation\cite{kudry1}. Even though the original form the method is based on a finite sum of the powers of a function including exponential function with base $e$, an arbitrary nonzero base $a$ is used in the following modifications of the method. Ege and Misirli\cite{ege1} use the modified form of the Kudryashov's method to derive the exact solutions of some fractional PDEs where the fractional derivatives are in the sense of Jumarie's modified Riemann-Liouville derivatives. Kaplan et al.\cite{kaplan1} implement the same technique to obtain the exact solutions of various equations as the solutions Rosenau-Kawahara equation are derived in \cite{tandogan1}. A recent study announces the exact solutions of some time fractional Klein-Gordon equations in the conformable derivative sense by using the modified Kudryashov method\cite{hosseini1}.

\section{Preliminaries and Essential Tools}
The conformable derivative of order $\alpha$ is defined as
\begin{equation}
T_{\alpha}(u)(t)=\lim\limits_{\tau\rightarrow 0}{\dfrac{u(t+\tau t^{1-\alpha})-u(t)}{\tau}}, \, t>0, \, \alpha \in (0,1].
\end{equation}
for a function $u:[0,\infty)\rightarrow \mathbb{R}$\cite{khalil1}. The conformable derivative satisfies the properties given in the following theorems.

\begin{theorem}
Assume that the order of the derivative $\alpha \in (0,1]$, and suppose that $u$ and $v$ are $\alpha$-differentiable for all positive $t$. Then,
\begin{itemize}
\item $T_{\alpha}(au+bv)=aT_{\alpha}(u)+bT_{\alpha}(v)$ 
\item $T_{\alpha}(t^p)=pt^{p-\alpha}, \forall p \in \mathbb{R}$
\item $T_{\alpha}(\lambda)=0$, for all constant function $u(t)=\lambda$
\item $T_{\alpha}(uv)=uT_{\alpha}(v)+vT_{\alpha}(u)$
\item $T_{\alpha}(\frac{u}{v})=\dfrac{uT_{\alpha}(v)-vT_{\alpha}(u)}{v^2}$
\item $T_{\alpha}(u)(t)=t^{1-\alpha}\frac{du}{dt}$
\end{itemize}
for $\forall a,b \in \mathbb{R}$\cite{atangana1,cenesiz1}.
\end{theorem}

\noindent
Some more properties covering the chain rule, Gronwall's inequality, some integration techniques, Laplace transform, Taylor series expansion and exponential function with respect to the conformable derivative are expressed in the work \cite{abdeljawad1}. 
\begin{theorem}
Let $u$ be an $\alpha$-differentiable function in conformable sense and differentiable and suppose that $v$ is also differentiable and defined in the range of $u$. Then,
\begin{equation}
T_{\alpha}(u\circ v)(t)=t^{1-\alpha}v^{\prime}(t)u^{\prime}(v(t))
\end{equation}
\end{theorem}

\section{Method of Solution}

\noindent
Consider a nonlinear PDE of the form
\begin{equation}
F(u,u_{t}^{\alpha},u_x,u_{t}^{2\alpha},u_{xx},...)=0 \label{gfpde}
\end{equation}
where $u=u(x,t)$ and $\alpha \in (0,1]$ stands for the order of the conformable derivative. The wave transformation
\begin{equation} 
u(x,t)=u(\xi), \xi = x-\frac{c}{\alpha}t^{\alpha} \label{wt}
\end{equation}
reduces the dimension of the (\ref{gfpde}) and generates and ordinary differential equation of the form
\begin{equation}
G(u,u{'},u{''},\ldots)=0 \label{gfode}
\end{equation}
where the prime ($'$) stands for the derivative of $u$ with respect to $\xi$. The variants of this transformation is used in the works \cite{hosseini1,eslami1}.

\noindent
Consider the equation (\ref{gfode}) has a solution of the form 
\begin{equation}
u(\xi)=\sum\limits_{i=0}^{n}{a_iQ^{n}(\xi)} \label{sol}
\end{equation}
with the necessary conditions $a_n\neq 0$ and all $a_i,0\leq i \leq n$ are constants and $$Q(\xi)=\dfrac{1}{1+dA^{\xi}}$$
solves 
\begin{equation}
Q^{'}(\xi)=Q(\xi)(Q(\xi)-1)\ln{A}
\label{qt}
\end{equation}
where $d$ and $A$ are nonzero constants with $A>0$ and $A\neq 1$. The determination of $n$ is completed by balancing the highest order derivative and the nonlinear terms in (\ref{gfode}). Thus, the solution (\ref{sol}) is substituted into (\ref{gfode}). Rearranging the resultant equation with respect to the powers of $Q(\xi)$ and solving the algebraic system derived from the coefficients of each power of $Q(\xi)$ gives $a_i,0\leq i \leq n$ and the relations among the other parameters if any. 

\section{Solution of the conformable fractional BBM equation}
\noindent
The conformable fractional BBM equation is defined as

\begin{equation}
u_t^{\alpha}+\epsilon u_x+\beta uu_x+\mu u_{xxt}^{\alpha}=0 \label{bbm0}
\end{equation}
where $\epsilon$, $\beta$, $\mu$ are parameters and $\alpha \in (0,1]$ denotes the conformable fractional derivative. The wave transformation (\ref{wt})reduces the dimension of the BBM equation to one and gives the ordinary differential equation
\begin{equation}
-cu'+\epsilon u' + \beta uu'-c\mu u'''=0 \label{bbm}
\end{equation}
where (') denotes derivative with respect to $\xi$. Integrating the equation (\ref{bbm}) gives

\begin{equation}
-cu+\epsilon u + \beta \dfrac{u^2}{2}-c\mu u''=K \label{bbm1}
\end{equation}
where $K$ is integral constant. Balancing the terms $u^2$ and $u''$ gives $n=2$. Thus, a solution in the form
\begin{equation}
u(\xi)=a_0+a_1Q(\xi)+a_2Q^2(\xi) \label{q}
\end{equation}
should be sought. Substituting the solution (\ref{q}) into (\ref{bbm1}) and using the property (\ref{qt}) sufficiently gives
\begin{equation}
\begin{aligned}
& \left( -6\,c\mu\,a_{{2}} \left( \ln  \left( A \right)  \right) ^{2}+\dfrac{1}{2}\,\beta\,{a_{{2}}}^{2} \right) Q ^{4}+ \left( -2\,c\mu\,a_{{1}} \left( \ln  \left( A \right) 
 \right) ^{2}+\beta\,a_{{1}}a_{{2}}+10\,c\mu\,a_{{2}} \left( \ln 
 \left( A \right)  \right) ^{2} \right)  Q  ^{3} \\
 &+ \left( \beta\,a_{{0}}a_{{2}}-ca_{{2}}+\epsilon\,a_{{2}}
+3\,c\mu\,a_{{1}} \left( \ln  \left( A \right)  \right) ^{2}-4\,c\mu\,
a_{{2}} \left( \ln  \left( A \right)  \right) ^{2}+\dfrac{1}{2}\,\beta\,{a_{{1}
}}^{2} \right)  Q ^{2} \\
&+ \left( 
\beta\,a_{{0}}a_{{1}}+\epsilon\,a_{{1}}-ca_{{1}}-c\mu\,a_{{1}} \left( 
\ln  \left( A \right)  \right) ^{2} \right)  Q +
\epsilon\,a_{{0}}+\dfrac{1}{2}\,\beta\,{a_{{0}}}^{2}-ca_{{0}}-K=0
 \end{aligned}
\end{equation}
in the rearranged form with respect to each power of $Q=Q(\xi)$. Equating the coefficients of each power of $Q$ and the constant term to zero gives the algebraic system

\begin{equation}
\begin{aligned}
-6\,c\mu\,a_{{2}} \left( \ln  \left( A \right)  \right) ^{2}+\dfrac{1}{2}\,
\beta\,{a_{{2}}}^{2}&=0\\
-2\,c\mu\,a_{{1}} \left( \ln  \left( A \right)  \right) ^{2}+\beta\,a_
{{1}}a_{{2}}+10\,c\mu\,a_{{2}} \left( \ln  \left( A \right)  \right) ^
{2}&=0\\
\beta\,a_{{0}}a_{{2}}-ca_{{2}}+\epsilon\,a_{{2}}+3\,c\mu\,a_{{1}}
 \left( \ln  \left( A \right)  \right) ^{2}-4\,c\mu\,a_{{2}} \left( 
\ln  \left( A \right)  \right) ^{2}+\dfrac{1}{2}\,\beta\,{a_{{1}}}^{2}&=0\\
\beta\,a_{{0}}a_{{1}}+\epsilon\,a_{{1}}-ca_{{1}}-c\mu\,a_{{1}} \left( 
\ln  \left( A \right)  \right) ^{2}&=0\\
\epsilon\,a_{{0}}+\dfrac{1}{2}\,\beta\,{a_{{0}}}^{2}-ca_{{0}}-K&=0 \label{algebraic1}
\end{aligned}
\end{equation}

\noindent
Solving this system for $a_0,a_1,a_2$ and $c$ gives
\begin{equation}
\begin{aligned}
a_0&= \dfrac{1}{\beta} \left( -\epsilon+{\frac {-\epsilon+B}{D}}+{\frac { \left( -\epsilon+B
 \right) \mu\, \left( \ln  \left( A \right)  \right) ^{2}}{D}}
 \right) \\
a_1&= \dfrac{-12\,\mu\, \left( \ln  \left( A \right)  \right) ^{2}}{\beta\left(-\epsilon+B\right)} \left( -2\,{
\frac {\epsilon\, \left( -\epsilon+B \right) }{D}}+{\epsilon}^{2}+2\,K
\beta \right)  \\
a_2&=12\,{\frac { \left( -\epsilon+B \right) \mu\, \left( \ln  \left( A
 \right)  \right) ^{2}}{D \beta}}\\
c&={\frac {-\epsilon+B}{D}}
\end{aligned}
\end{equation}
where
\begin{equation}
\begin{aligned}
B&=  \sqrt {-2\,K\beta+{\mu}^{2}
 \left( \ln  \left( A \right)  \right) ^{4}{\epsilon}^{2}+2\,{\mu}^{2}
 \left( \ln  \left( A \right)  \right) ^{4}K\beta}\\
D&=-1+{\mu}^{2} \left( \ln  \left( A \right)  \right) ^{4}
\end{aligned}
\end{equation}
Thus, the solution is expressed in an explicit form as
\begin{equation}
\begin{aligned}
u(\xi)&= \dfrac{1}{\beta} \left( -\epsilon+{\frac {-\epsilon+B}{D}}+{\frac { \left( -\epsilon+B
 \right) \mu\, \left( \ln  \left( A \right)  \right) ^{2}}{D}}
 \right) \\
 &-\dfrac{12\,\mu\, \left( \ln  \left( A \right)  \right) 
^{2}}{\left( -\epsilon+B \right) {\beta} } \left( -2\,{\frac {\epsilon\, \left( -\epsilon+B \right) }{D}}+{
\epsilon}^{2}+2\,K\beta \right) \dfrac{1}{1+d{A}^{x-c\dfrac{t^{\alpha}}{\alpha}}}  \\
&+12\,{\frac { \left( -
\epsilon+B \right) \mu\, \left( \ln  \left( A \right)  \right) ^{2}}{D \beta}\dfrac{1}{ \left( 1+d{A}^{x-c\dfrac{t^{\alpha}}{\alpha}} \right) ^{2}}}
\end{aligned}
\end{equation}
Returning the original independent variables $x$ and $t$ gives the solution as
\begin{equation}
\begin{aligned}
u_1(x,t)&= \dfrac{1}{\beta} \left( -\epsilon+{\frac {-\epsilon+B}{D}}+{\frac { \left( -\epsilon+B
 \right) \mu\, \left( \ln  \left( A \right)  \right) ^{2}}{D}}
 \right) \\
 &-\dfrac{12\,\mu\, \left( \ln  \left( A \right)  \right) 
^{2}}{\left( -\epsilon+B \right) {\beta} } \left( -2\,{\frac {\epsilon\, \left( -\epsilon+B \right) }{D}}+{
\epsilon}^{2}+2\,K\beta \right) \dfrac{1}{1+d{A}^{x-c\dfrac{t^{\alpha}}{\alpha}}}  \\
&+12\,{\frac { \left( -
\epsilon+B \right) \mu\, \left( \ln  \left( A \right)  \right) ^{2}}{D \beta}\dfrac{1}{ \left( 1+d{A}^{x-c\dfrac{t^{\alpha}}{\alpha}} \right) ^{2}}}
\end{aligned} \label{u1}
\end{equation}
where $c$ is given above.

\noindent
A second solution can be derived
\begin{equation}
\begin{aligned}
a_0&=\dfrac{1}{\beta}\left( -\epsilon-{\dfrac{\epsilon+B}{D}}-{\frac { \left( \epsilon+B
 \right) \mu\, \left( \ln  \left( A \right)  \right) ^{2}}{D}}
 \right) \\
a_1&= \dfrac{12\,\mu\, \left( \ln  \left( A \right)  \right) ^{2}}{\left( \epsilon+B \right) {\beta}} \left( 2\,{\frac 
{\epsilon\, \left( \epsilon+B \right) }{D}}+{\epsilon}^{2}+2\,K\beta
 \right)  \\
a_2&=-12\,{\frac { \left( \epsilon+B \right) \mu\, \left( \ln  \left( A
 \right)  \right) ^{2}}{  D \beta}}\\
c&=-{\frac {\epsilon+B}{D}}
\end{aligned}
\end{equation}
to the system (\ref{algebraic1}). These coefficients generate the solution 
\begin{equation}
\begin{aligned}
u(\xi)&=  \dfrac{1}{\beta}\left( -\epsilon-{\dfrac{\epsilon+B}{D}}-{\frac { \left( \epsilon+B
 \right) \mu\, \left( \ln  \left( A \right)  \right) ^{2}}{D}}
 \right) \\
&+\dfrac{12\,\mu\, \left( \ln  \left( A \right)  \right) ^{2}}{\left( \epsilon+B \right) {\beta}} \left( 2\,{\frac 
{\epsilon\, \left( \epsilon+B \right) }{D}}+{\epsilon}^{2}+2\,K\beta
 \right) \dfrac{1}{1+d{A}^{\xi}}\\
&-12\,{\frac { \left( \epsilon+
B \right) \mu\, \left( \ln  \left( A \right)  \right) ^{2}}{ D
 \beta\, \left( 1+d{A}^{\xi} \right) ^{2}}}
\end{aligned}
\end{equation}
Using the original variables converts the solution  to
\begin{equation}
\begin{aligned}
u_2(x,t)&=  \dfrac{1}{\beta}\left( -\epsilon-{\dfrac{\epsilon+B}{D}}-{\frac { \left( \epsilon+B
 \right) \mu\, \left( \ln  \left( A \right)  \right) ^{2}}{D}}
 \right) \\
&+\dfrac{12\,\mu\, \left( \ln  \left( A \right)  \right) ^{2}}{\left( \epsilon+B \right) {\beta}} \left( 2\,{\frac 
{\epsilon\, \left( \epsilon+B \right) }{D}}+{\epsilon}^{2}+2\,K\beta
 \right) \dfrac{1}{1+d{A}^{x-c\dfrac{t^{\alpha}}{\alpha}}}\\
&-12\,{\frac { \left( \epsilon+
B \right) \mu\, \left( \ln  \left( A \right)  \right) ^{2}}{ D
 \beta\, \left( 1+d{A}^{x-c\dfrac{t^{\alpha}}{\alpha}} \right) ^{2}}} 
\end{aligned}\label{u2}
\end{equation}

\begin{figure}[hp]
\centering
   \subfigure[$\alpha =0.25$]{
   \includegraphics[scale =0.35] {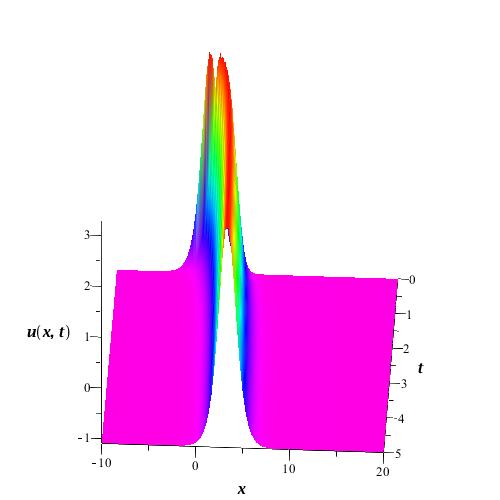}
   \label{rlw1}
 }
   \subfigure[$\alpha =0.50$]{
   \includegraphics[scale =0.35] {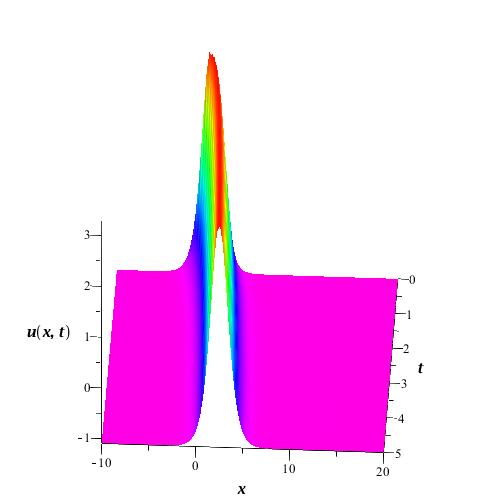}
   \label{rlw2}
 }
    \subfigure[$\alpha =0.80$]{
   \includegraphics[scale =0.35] {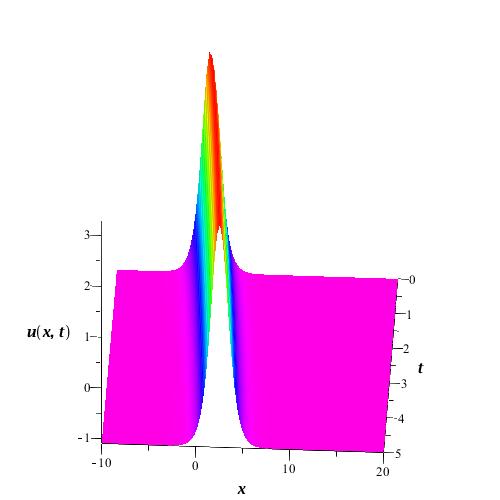}
   \label{rlw3}
 }
    \subfigure[$\alpha =1$]{
   \includegraphics[scale =0.35] {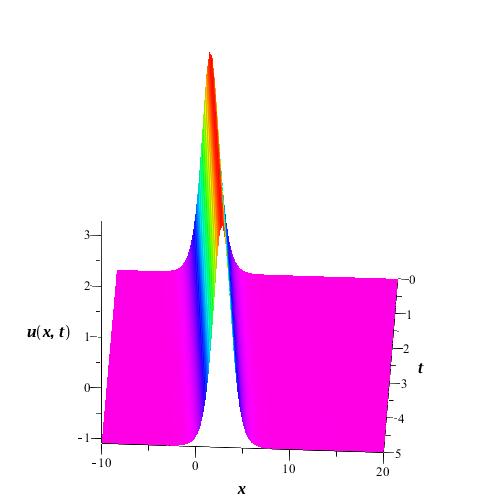}
   \label{rlw4}
 }
 
\caption{The solution $u_1(x,t)$ for the parameters $\epsilon =1/5$, $\beta =1$, $\mu =-1$, $A=1/5$, $d=1$, $K=1$}
\end{figure}

\noindent
The illustrations of the solution (\ref{u1}) for various values of $\alpha$ are depicted in Fig \ref{rlw1}-\ref{rlw4}.

\section{Solution of the conformable fractional symmetric BBM equation}
\begin{figure}[hp]
\centering
   \subfigure[$\alpha =0.25$]{
   \includegraphics[scale =0.35] {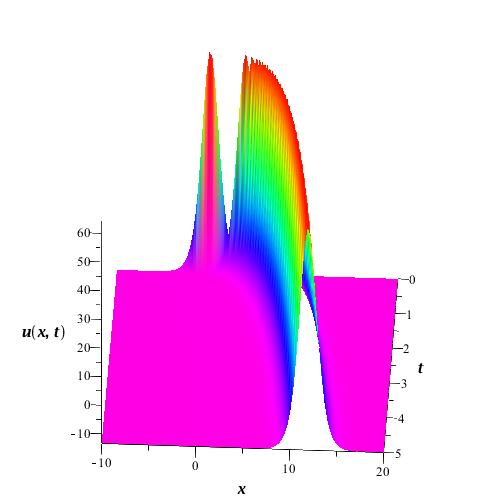}
   \label{srlw1}
 }
   \subfigure[$\alpha =0.50$]{
   \includegraphics[scale =0.35] {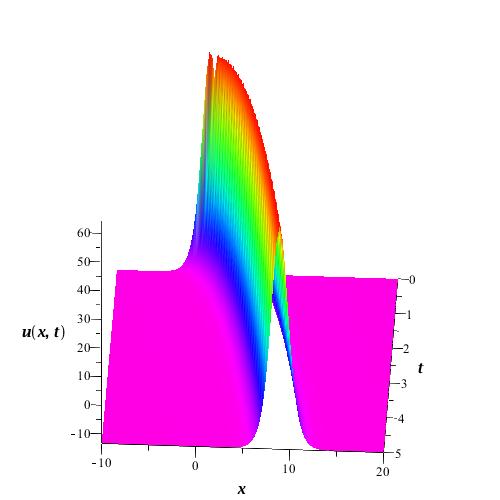}
   \label{srlw2}
 }
    \subfigure[$\alpha =0.80$]{
   \includegraphics[scale =0.35] {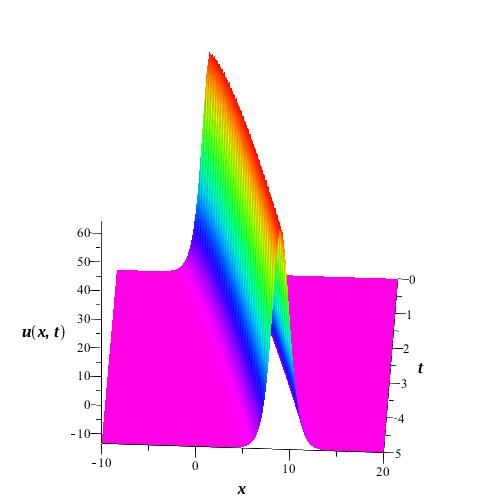}
   \label{srlw3}
 }
    \subfigure[$\alpha =1$]{
   \includegraphics[scale =0.35] {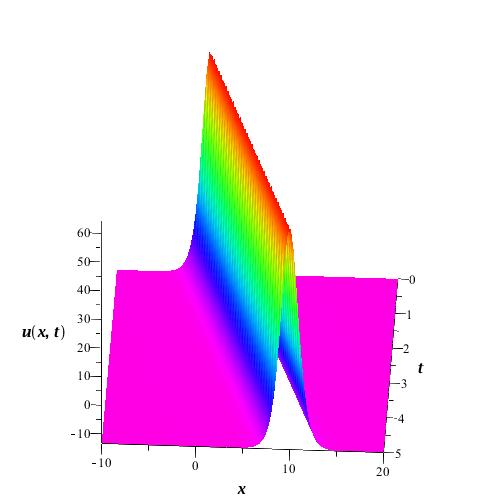}
   \label{srlw4}
 }
 
\caption{The solution $u_3(x,t)$ for the parameters $\delta =1/5$, $\beta =1$, $\mu =-1$, $A=1/5$, $d=1$, $K=0$,$c=2$}
\end{figure}

The conformable fractional symmetric BBM equation of the form
\begin{equation}
u_{tt}^{2\alpha}+\beta u_{xx}+\delta uu_{xt}^{\alpha}+\delta u_x u_{t}^{\alpha}+\mu u_{xxtt}^{2\alpha}=0
\end{equation}
can be reduced to the ordinary differential equation
\begin{equation}
\left(c^2+\beta\right)u^{''}-c\delta\left(uu^{''}+\left(u^{'}\right)^2\right)+c^2\mu u^{''''}=0
\end{equation}
by using the wave transformation (\ref{wt}). Integrating this ordinary differential equation results
\begin{equation}
\left(c^2+\beta\right)u^{'}-c\delta uu^{'}+c^2\mu u^{'''}=K \label{sbbmo}
\end{equation}
where $K$ is the constant of integration. The balance between $uu^{'}$ and $u^{'''}$ leads $n=2$. However, the substitution of the solution of the form
\begin{equation}
u(\xi)=a_0+a_1Q(\xi)+a_2Q^2(\xi)
\end{equation}
into (\ref{sbbmo}) gives 
\begin{equation}
\begin{aligned}
&\left( -2\,c\delta\,{a_{{2}}}^{2}\ln  \left( A \right) +24\,{c}^{2}
\mu\,a_{{2}} \left( \ln  \left( A \right)  \right) ^{3} \right) 
 \left( Q \left( \xi \right)  \right) ^{5}\\&
 + \left( -54\,{c}^{2}\mu\,a_
{{2}} \left( \ln  \left( A \right)  \right) ^{3}+6\,{c}^{2}\mu\,a_{{1}
} \left( \ln  \left( A \right)  \right) ^{3}+2\,c\delta\,{a_{{2}}}^{2}
\ln  \left( A \right) -3\,c\delta\,a_{{1}}\ln  \left( A \right) a_{{2}
} \right)  \left( Q \left( \xi \right)  \right) ^{4}\\
&+ \left[ 2\,\beta
\,a_{{2}}\ln  \left( A \right) -12\,{c}^{2}\mu\,a_{{1}} \left( \ln 
 \left( A \right)  \right) ^{3}+38\,{c}^{2}\mu\,a_{{2}} \left( \ln 
 \left( A \right)  \right) ^{3}+2\,{c}^{2}a_{{2}}\ln  \left( A
 \right) -c\delta\,{a_{{1}}}^{2}\ln  \left( A \right) \right. \\
 &\left. -2\,c\delta\,a_{
{2}}\ln  \left( A \right) a_{{0}}+3\,c\delta\,a_{{1}}\ln  \left( A
 \right) a_{{2}} \right]  \left( Q \left( \xi \right)  \right) ^{3}\\
 &+
 \left( {c}^{2}a_{{1}}\ln  \left( A \right) -2\,\beta\,a_{{2}}\ln 
 \left( A \right) -8\,{c}^{2}\mu\,a_{{2}} \left( \ln  \left( A
 \right)  \right) ^{3}-c\delta\,a_{{1}}\ln  \left( A \right) a_{{0}}+7
\,{c}^{2}\mu\,a_{{1}} \left( \ln  \left( A \right)  \right) ^{3}\right.\\
&\left.+\beta
\,a_{{1}}\ln  \left( A \right) +2\,c\delta\,a_{{2}}\ln  \left( A
 \right) a_{{0}}-2\,{c}^{2}a_{{2}}\ln  \left( A \right) +c\delta\,{a_{
{1}}}^{2}\ln  \left( A \right)  \right)  \left( Q \left( \xi \right) 
 \right) ^{2}\\
 &+ \left( -\beta\,a_{{1}}\ln  \left( A \right) -{c}^{2}a_{
{1}}\ln  \left( A \right) +c\delta\,a_{{1}}\ln  \left( A \right) a_{{0
}}-{c}^{2}\mu\,a_{{1}} \left( \ln  \left( A \right)  \right) ^{3}
 \right) Q \left( \xi \right) -K=0
 \end{aligned}\label{1}
\end{equation}
where $K$ is the integral constant. The solution of the system of the algebraic equations constructed by using the coefficients of each power of the $Q(\xi)$ and the constant term in (\ref{1}) gives
\begin{equation}
\begin{aligned}
a_{0}&={\frac {{c}^{2}\mu\, \left( \ln  \left( A \right)  \right) ^{2}+\beta+
{c}^{2}}{c\delta}} \\
a_1&=-12\,{\frac {c\mu\, \left( \ln  \left( A \right)  \right) ^{2}}{\delta
}} \\
a_2&=12\,{\frac {c\mu\, \left( \ln  \left( A \right)  \right) ^{2}}{\delta}
}
\end{aligned}
\end{equation}
for arbitrary nonzero $c$ and $K=0$. Using $\{a_0,a_1,a_2\}$, the solution can be constructed as
\begin{equation}
\begin{aligned}
u(\xi)={\frac {{c}^{2}\mu\, \left( \ln  \left( A \right)  \right) ^{2}+\beta+
{c}^{2}}{c\delta}} -12\,{\frac {c\mu\, \left( \ln  \left( A \right)  \right) ^{2}}{\delta
}} \dfrac{1}{1+dA^{\xi}}+12\,{\frac {c\mu\, \left( \ln  \left( A \right)  \right) ^{2}}{\delta}\dfrac{1}{\left(1+dA^{\xi}\right)^2}}
\end{aligned}
\end{equation}
Returning the original variables gives the solution as

\begin{equation}
\begin{aligned}
u_3(x,t)&={\frac {{c}^{2}\mu\, \left( \ln  \left( A \right)  \right) ^{2}+\beta+
{c}^{2}}{c\delta}} -12\,{\frac {c\mu\, \left( \ln  \left( A \right)  \right) ^{2}}{\delta
}} \dfrac{1}{1+dA^{x-c\dfrac{t^{\alpha}}{\alpha}}} \\
&+12\,{\frac {c\mu\, \left( \ln  \left( A \right)  \right) ^{2}}{\delta}\dfrac{1}{\left(1+dA^{x-c\dfrac{t^{\alpha}}{\alpha}}\right)^2}}
\end{aligned}\label{u3}
\end{equation}
The illustrations of the solution $u_3(x,t)$(\ref{u3}) of the conformable time fractional symmetric BBM equation for various values of $\alpha$ are depicted in Fig \ref{srlw1}-\ref{srlw4}.

\section{Solution of the conformable fractional EW equation}
\begin{figure}[hp]
\centering
   \subfigure[$\alpha =0.25$]{
   \includegraphics[scale =0.35] {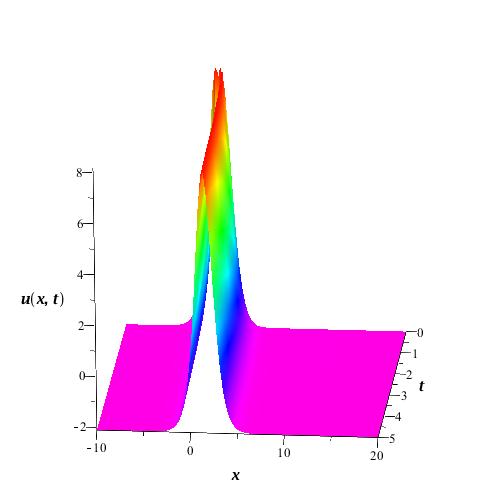}
   \label{ew1}
 }
   \subfigure[$\alpha =0.50$]{
   \includegraphics[scale =0.35] {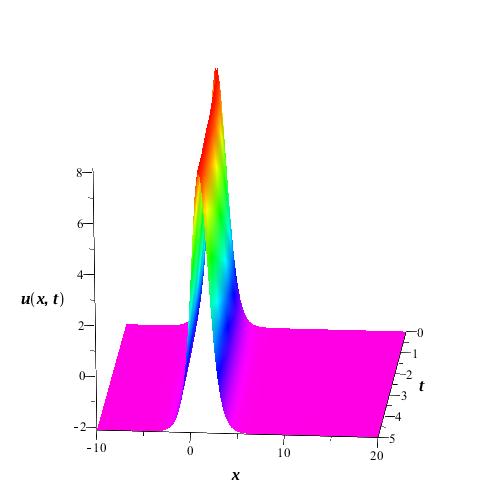}
   \label{ew2}
 }
    \subfigure[$\alpha =0.80$]{
   \includegraphics[scale =0.35] {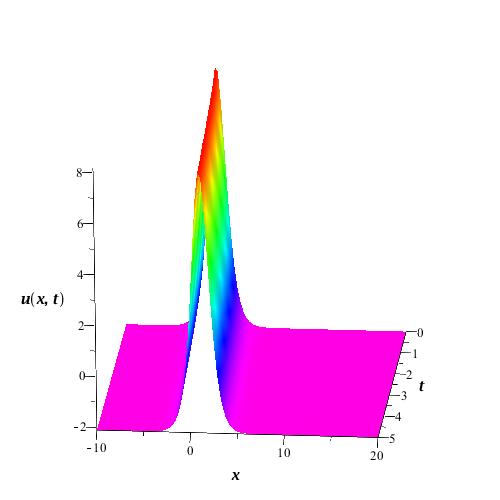}
   \label{ew3}
 }
    \subfigure[$\alpha =1$]{
   \includegraphics[scale =0.35] {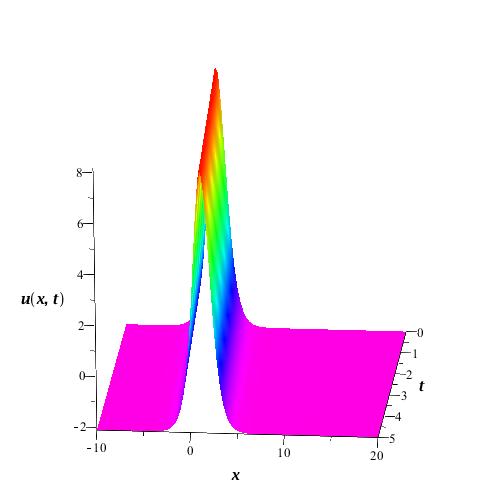}
   \label{ew4}
 }
 
\caption{The solution $u_4(x,t)$ for the parameters $\epsilon =1/5$, $\delta =1$, $A=1/5$, $d=1$, $K=1$}
\end{figure}

The conformable time fractional EW equation is defined as
\begin{equation}
u_t^{\alpha}+\epsilon u u_x -\delta u_{xxt}^{\alpha}=0
\label{ew}
\end{equation}
where $\epsilon$ and $\delta$ are nonzero parameters. The wave transformation (\ref{wt} reduces the conformable time fractional EW equation to
\begin{equation}
-cu^{'}+\dfrac{\epsilon}{2}(u^2)^{'}+\delta c u^{'''}=0
\end{equation}
where $^{'}$ stands for the derivative with respect to $\xi$. When the last equation is integrated, it becomes
\begin{equation}
-cu+\dfrac{\epsilon}{2}(u^2)+\delta c u^{''}=K \label{ewo}
\end{equation}
where $K$ is integration constant. Balancing $u^2$ and $u^{''}$ gives $N=2$. Thus, a solution in the form
\begin{equation}
u(\xi)=a_0+a_1Q(\xi)+a_2Q^2(\xi)
\end{equation}
and satisfying the conditions given in the previous sections can be investigated. Substituting the solution into (\ref{ewo}) leads
\begin{equation}
\begin{aligned}
 &\left( 1/2\,\epsilon\,{a_{{2}}}^{2}+6\,\delta\,ca_{{2}} \left( \ln 
 \left( A \right)  \right) ^{2} \right)  \left( Q \left( \xi \right) 
 \right) ^{4} \\
 &+ \left( 2\,\delta\,ca_{{1}} \left( \ln  \left( A
 \right)  \right) ^{2}-10\,\delta\,ca_{{2}} \left( \ln  \left( A
 \right)  \right) ^{2}+\epsilon\,a_{{1}}a_{{2}} \right)  \left( Q
 \left( \xi \right)  \right) ^{3}\\
 &+ \left( 1/2\,\epsilon\,{a_{{1}}}^{2}
+\epsilon\,a_{{0}}a_{{2}}-3\,\delta\,ca_{{1}} \left( \ln  \left( A
 \right)  \right) ^{2}+4\,\delta\,ca_{{2}} \left( \ln  \left( A
 \right)  \right) ^{2}-ca_{{2}} \right)  \left( Q \left( \xi \right) 
 \right) ^{2}\\
 &+ \left( -ca_{{1}}+\epsilon\,a_{{0}}a_{{1}}+\delta\,ca_{{
1}} \left( \ln  \left( A \right)  \right) ^{2} \right) Q \left( \xi
 \right) +1/2\,\epsilon\,{a_{{0}}}^{2}-K-ca_{{0}}=0
\end{aligned}
\end{equation}
in the rearranged form with respect to the powers of $Q(\xi)$. Equating the coefficients of the powers of the $Q(\xi)$ to zero gives the algebraic system
\begin{equation}
\begin{aligned}
1/2\,\epsilon\,{a_{{2}}}^{2}+6\,\delta\,ca_{{2}} \left( \ln  \left( A
 \right)  \right) ^{2}&=0 \\
2\,\delta\,ca_{{1}} \left( \ln  \left( A \right)  \right) ^{2}-10\,
\delta\,ca_{{2}} \left( \ln  \left( A \right)  \right) ^{2}+\epsilon\,
a_{{1}}a_{{2}}&=0 \\
1/2\,\epsilon\,{a_{{1}}}^{2}+\epsilon\,a_{{0}}a_{{2}}-3\,\delta\,ca_{{
1}} \left( \ln  \left( A \right)  \right) ^{2}+4\,\delta\,ca_{{2}}
 \left( \ln  \left( A \right)  \right) ^{2}-ca_{{2}}&=0 \\
-ca_{{1}}+\epsilon\,a_{{0}}a_{{1}}+\delta\,ca_{{1}} \left( \ln 
 \left( A \right)  \right) ^{2}&=0 \\
1/2\,\epsilon\,{a_{{0}}}^{2}-K-ca_{{0}}&=0 \label{ewalg}
\end{aligned}
\end{equation}
and the solution of this system for $a_0,a_1,a_2,c$ gives
\begin{equation}
\begin{aligned}
a_0&= -{\frac {\sqrt {2}  D  \left( -1+\delta\, \left( \ln 
 \left( A \right)  \right) ^{2} \right) }{\epsilon}}\\
a_1&= 12\,{\frac {K\delta\, \left( \ln  \left( A \right)  \right) ^{2}\sqrt 
{2}}{  D  \left( -1+{\delta}^{2} \left( \ln  \left( A
 \right)  \right) ^{4} \right) }}\\
 a_2&= -12\,{\frac {\delta\,\sqrt {2}  D \left( \ln  \left( A
 \right)  \right) ^{2}}{\epsilon}}\\
 c&=\sqrt{2}D
\end{aligned}
\end{equation}
where $D=\sqrt {{\frac {K\epsilon}{-1+{\delta}^{2} \left( \ln  \left( A
 \right)  \right) ^{4}}}}$. Thus, the solution for the independent variable $\xi$ can be written as
\begin{equation}
\begin{aligned}
u(\xi)&=-{\frac {\sqrt {2}  D  \left( -1+\delta\, \left( \ln 
 \left( A \right)  \right) ^{2} \right) }{\epsilon}}+ 12\,{\frac {K\delta\, \left( \ln  \left( A \right)  \right) ^{2}\sqrt 
{2}}{  D  \left( -1+{\delta}^{2} \left( \ln  \left( A
 \right)  \right) ^{4} \right) }} \dfrac{1}{\left(1+dA^{\xi}\right)} \\
 &-12\,{\frac {\delta\,\sqrt {2}  D \left( \ln  \left( A
 \right)  \right) ^{2}}{\epsilon}} \dfrac{1}{\left(1+dA^{\xi}\right)^2}
\end{aligned}
\end{equation}
Returning to the original variables gives the solution in terms of $x$ and $t$ as
\begin{equation}
\begin{aligned}
u_4(x,t)&=-{\frac {\sqrt {2}  D  \left( -1+\delta\, \left( \ln 
 \left( A \right)  \right) ^{2} \right) }{\epsilon}}+ 12\,{\frac {K\delta\, \left( \ln  \left( A \right)  \right) ^{2}\sqrt 
{2}}{  D  \left( -1+{\delta}^{2} \left( \ln  \left( A
 \right)  \right) ^{4} \right) }} \dfrac{1}{\left(1+dA^{x-\sqrt{2}D\dfrac{t^{\alpha}}{\alpha}}\right)} \\
 &-12\,{\frac {\delta\,\sqrt {2}  D \left( \ln  \left( A
 \right)  \right) ^{2}}{\epsilon}} \dfrac{1}{\left(1+dA^{x-\sqrt{2}D\dfrac{t^{\alpha}}{\alpha}}\right)^2}
\end{aligned}
\end{equation}
\noindent
The other solution
\begin{equation}
\begin{aligned}
a_0&= {\frac {\sqrt {2} D   \left( -1+\delta\, \left( \ln 
 \left( A \right)  \right) ^{2} \right) }{\epsilon}}\\
a_1&= -12\,{\frac {K\delta\, \left( \ln  \left( A \right)  \right) ^{2}
\sqrt {2}}{  D  \left( -1+{\delta}^{2} \left( \ln 
 \left( A \right)  \right) ^{4} \right) }}\\
 a_2&= 12\,{\frac {\delta\,\sqrt {2}  D   \left( \ln  \left( A
 \right)  \right) ^{2}}{\epsilon}}\\
 c&=-\sqrt{2}D
\end{aligned}
\end{equation}
where $D=\sqrt {{\frac {K\epsilon}{-1+{\delta}^{2} \left( \ln  \left( A
 \right)  \right) ^{4}}}}$.
of the algebraic system (\ref{ewalg}. This coefficients gives the solution for $\xi$ as
\begin{equation}
\begin{aligned}
u(\xi)&={\frac {\sqrt {2}  D  \left( -1+\delta\, \left( \ln 
 \left( A \right)  \right) ^{2} \right) }{\epsilon}}- 12\,{\frac {K\delta\, \left( \ln  \left( A \right)  \right) ^{2}\sqrt 
{2}}{  D  \left( -1+{\delta}^{2} \left( \ln  \left( A
 \right)  \right) ^{4} \right) }} \dfrac{1}{\left(1+dA^{\xi}\right)} \\
 &12\,{\frac {\delta\,\sqrt {2}  D \left( \ln  \left( A
 \right)  \right) ^{2}}{\epsilon}} \dfrac{1}{\left(1+dA^{\xi}\right)^2}
\end{aligned}
\end{equation}  
Changing the variable $\xi$ to $x$ and $t$ gives
\begin{equation}
\begin{aligned}
u_5(x,t)&={\frac {\sqrt {2}  D  \left( -1+\delta\, \left( \ln 
 \left( A \right)  \right) ^{2} \right) }{\epsilon}}- 12\,{\frac {K\delta\, \left( \ln  \left( A \right)  \right) ^{2}\sqrt 
{2}}{  D  \left( -1+{\delta}^{2} \left( \ln  \left( A
 \right)  \right) ^{4} \right) }} \dfrac{1}{\left(1+dA^{x+\sqrt{2}D\dfrac{t^{\alpha}}{\alpha}}\right)} \\
 &+12\,{\frac {\delta\,\sqrt {2}  D \left( \ln  \left( A
 \right)  \right) ^{2}}{\epsilon}} \dfrac{1}{\left(1+dA^{x+\sqrt{2}D\dfrac{t^{\alpha}}{\alpha}}\right)^2}
\end{aligned}
\end{equation}
The solution $u_4(x,t)$ is illustrated for various values of the conformable fractional derivative in Fig \ref{ew1}-\ref{ew3}.
\section{Conclusion}
\noindent
Exact solutions of some conformable time fractional PDEs in the BBM family are constructed by implementing the Kudryashov method. The compatible wave transformation reduces the considered PDE to an integer-ordered ordinary differential equation. The balance between the nonlinear and the highest ordered derivative may enable to be existence of the solutions expressed in the finite series of a particular function satisfying a first order differential equation. Substituting this predicted solution into the resulted ODE gives the relation between the coefficients of the terms of the finite series if exist. 

\noindent
The successful implementations of the method yield such solutions for the conformable time fractional BBM, the symmetric BBM and the EW equations. Those solutions are illustrated for some particular choices of the parameters.

\end{document}